\documentclass[10pt,twocolumn,a4paper]{ieeetran}
\usepackage{amsmath}
\usepackage{graphicx}
\usepackage{multirow}

\newcommand{\figuramedia}[3]
{
\begin{figure}
  \centering
 \includegraphics[width=8cm]{#1}
  \caption{#2}\label{#3}
\end{figure}
}

\newcommand{\figura}[3]
{
\begin{figure}
  \centering
 \includegraphics[width=8cm]{#1}
  \caption{#2}\label{#3}
\end{figure}
}
\title{On the Throughput Allocation for Proportional Fairness in Multirate IEEE 802.11 DCF under General Load Conditions}
\author{\authorblockN{F. Daneshgaran, M. Laddomada, F. Mesiti, and M.
Mondin
\thanks{F. Daneshgaran is with ECE Dept., California State University,
Los Angeles, USA.}
\thanks{M. Laddomada, F. Mesiti, and M. Mondin are with DELEN, Politecnico di Torino, Italy.}}
\authorblockA{}}
\begin{document}
\maketitle
\begin{abstract}
This paper presents a modified proportional fairness (PF)
criterion suitable for mitigating the \textit{rate anomaly}
problem of multirate IEEE 802.11 Wireless LANs employing the
mandatory Distributed Coordination Function (DCF) option. Compared
to the widely adopted assumption of saturated network, the
proposed criterion can be applied to general networks whereby the
contending stations are characterized by specific packet arrival
rates, $\lambda_s$, and transmission rates $R_d^{s}$.

The throughput allocation resulting from the proposed algorithm is
able to greatly increase the aggregate throughput of the DCF while
ensuring fairness levels among the stations of the same order of
the ones available with the classical PF criterion. Put simply,
each station is allocated a throughput that depends on a suitable
normalization of its packet rate, which, to some extent, measures
the frequency by which the station tries to gain access to the
channel. Simulation results are presented for some sample
scenarios, confirming the effectiveness of the proposed criterion.
\end{abstract}
\section{Introduction}
Consider the IEEE802.11 Medium Access Control (MAC) layer~\cite{standard_DCF_MAC} employing the DCF
based on the Carrier Sense Multiple Access Collision Avoidance CSMA/CA access method. The scenario
envisaged in this work considers $N$ contending stations; each station generates data packets with
constant rate $\lambda_s$ by employing a bit rate, $R_d^{s}$, which depends on the channel quality
experienced. In this scenario, it is known that the DCF is affected by the so-called
\textit{performance anomaly} problem~\cite{heusse}: in multirate networks the aggregate throughput
is strongly influenced by that of the slowest contending station.

After the landmark work by Bianchi~\cite{Bianchi}, who provided an
analysis of the saturation throughput of the basic 802.11 protocol
assuming a two dimensional Markov model at the MAC layer, many
papers have addressed almost any facet of the behaviour of DCF in
a variety of traffic loads and channel transmission conditions
(see \cite{daneshgaran_wcnc08}-\cite{daneshgaran_multirate} and
references therein). A current topic of interest in connection to
DCF regards the allocation of throughput in order to mitigate the
performance anomaly, while ensuring fairness among multirate
stations.

With this background, let us provide a quick survey of the recent
literature related to the problem addressed here. Paper
\cite{banchs} proposes a proportional fairness throughput
allocation criterion for multirate and saturated IEEE 802.11 DCF
by focusing on the 802.11e standard. In papers
\cite{TinnirelloI}-\cite{joshi2} the authors propose novel
fairness criteria, which fall within the class of the time-based
fairness criterion. Time-based fairness guarantees equal
time-share of the channel occupancy irrespective of the station
bit rate. Finally, paper \cite{babu_1} investigates the fairness
problem in 802.11 multirate networks from a theoretical point of
view.

A common hypothesis employed in the literature regards the
saturation assumption. However, real networks are different in
many respects. Traffic is mostly non-saturated, different stations
usually operate with different loads, i.e., they have different
packet rates, while the transmitting bit rate can also differ
among the contending stations. In all these situations the common
hypothesis, widely employed in the literature, that all the
contending stations have the same probability of transmitting in a
randomly chosen time slot, does not hold anymore. The aim of this
paper is to present a proportional fairness criterion under much
more realistic scenarios, especially when the packet rates among
the station are different. As a starting point for the derivations
that follow, we consider the bi-dimensional Markov model proposed
in the companion paper \cite{daneshgaran_multirate}, and present
the necessary modifications in order to deal with the problem at
hand.

The rest of the paper is organized as follows.
Section~\ref{sec:multirate_model} provides the necessary
modifications to the Markov model proposed in
\cite{daneshgaran_multirate}. For conciseness, we invite the
interested reader to refer to \cite{daneshgaran_multirate} for
further details about the bi-dimensional Markov model. The novel
proportional fairness criterion is presented in
Section~\ref{sec:optimization}. Finally, simulation results of
sample network scenarios are discussed in
Section~\ref{SimulationResults_Section}.
\section{Extension to the Markovian Model Characterizing the DCF}
\label{sec:multirate_model}
%
%
In a companion paper \cite{daneshgaran_multirate} (see also
\cite{daneshgaran_wcnc08}), we derived a bi-dimensional Markov
model for characterizing the behavior of the DCF under a variety
of real traffic conditions, both non-saturated and saturated, with
packet queues of small sizes, multirate setting, and considered
the IEEE 802.11b protocol with the basic 2-way handshaking
mechanism. As a starting point for the derivations which follow,
we adopt the bi-dimensional model proposed in
~\cite{daneshgaran_multirate}, appropriately modified in order to
account for the following scenario. In the investigated network,
each station employs a specific bit rate, $R_d^{(s)}$, a different
transmission packet rate, $\lambda_s$, transmits packets with size
$PL^{(s)}$, and it employs a minimum contention window with size
$W_0^{(s)}$, which can differ from the one specified in the IEEE
802.11 standard \cite{standard_DCF_MAC} (this is required for
optimizing the aggregate throughput while guaranteeing fairness
among the contending stations). For the sake of greatly
simplifying the evaluation of the expected time slots required by
the theoretical derivations that follow, we consider $N_c \leq N$
classes of channel occupancy durations. This assumption relies on
the observation that in actual networks some stations might
transmit data frames presenting the same channel occupancy. 

The scenario at hand requires a number of modifications to the previous theoretical results derived
in \cite{daneshgaran_multirate}. First of all, given the payload lengths and the data rates of the
$N$ stations, the $N_c$ duration-classes are arranged in order of decreasing durations identified
by the index $d \in \{1, \cdots, N_c\}$, whereby $d=1$ identifies the slowest class. Notice that in
our setup a station is labelled as fast if it has a short channel occupancy. Furthermore, each
station is identified by an index $s \in \{1, \cdots, N\}$, and it belongs to a unique
duration-class. In order to identify the class of a station $s$, we define $N_c$ subsets $n(d)$,
each of them containing the indexes of the $L_d=|n(d)|$ stations within $n(d)$, with $L_d \leq N,
\forall d$ and $\sum_{d=1}^{N_c}L_d = N$. As an instance, $n(3) = \{1,5,8\}$ means that stations 1,
5, and 8 belong to the third duration-class identified by $d=3$, and $L_d=3$.

With this setup, the probability that the $s$-th station starts a
transmission in a randomly chosen time slot is identified by
$\tau_s$, and it can be obtained by solving the bidimensional
Markov chain for the contention model of the $s$-th station
\cite{daneshgaran_multirate}:
\begin{equation}\label{eq:tau_s}\small
\tau_{s} =
\frac{2(1-b_I^{(s)})(1-2P_{eq}^{(s)})}{(W_0^{(s)}+1)(1-2P_{eq}^{(s)})
+ W_0^{(s)}P_{eq}^{(s)}[1-(2P_{eq}^{(s)})^m]}
\end{equation}
whereby $b_I^{(s)}$ is the stationary probability to be in an idle
state modelling unloaded conditions\footnote{Briefly, this state
models the situations when, immediately after a successful
transmission, the queue of the transmitting station is empty, or
the station is in an idle state with an empty queue until a new
packet arrives in the queue.}, $W_0^{(s)}$ is the minimum
contention window size of the $s$-th station, and $P^{(s)}_{eq}$
is the probability of equivalent failed transmission defined as
%
$P^{(s)}_{eq}=1-(1-P_{e}^{(s)})(1-P_{col}^{(s)})=P^{(s)}_{col}+P^{(s)}_e-P^{(s)}_e\cdot
P^{(s)}_{col}$,
%
%
whereby $P^{(s)}_{col}$ and $P^{(s)}_e$ are, respectively, the
collision and the packet error probabilities related to the $s$-th
station. Given $\tau_{s}$ in (\ref{eq:tau_s}), we can evaluate the
aggregate throughput $S$ as follows:
\begin{equation}\label{eq:throughput_aggr}
  S = \sum_{s=1}^{N} S_s= \sum_{s=1}^{N}\frac{1}{T_{av}}P_s^{(s)} \cdot (1 - P_e^{(s)}) \cdot PL^{(s)}
\end{equation}
whereby $T_{av}$ is the expected time per slot, $PL^{(s)}$ is the
packet size of the $s$-th station, and $P_s^{(s)}$ is the
probability of successful packet transmission of the $s$-th
station:
\begin{equation} \label{eq:pSucc_s}
P_s^{(s)}  =  \tau_{s}\cdot \prod_{\substack{j = 1 \\ j \neq
s}}^{N} (1 - \tau_{j})
\end{equation}
The expected time per slot, $T_{av}$, can be evaluated by
weighting the times spent by a station in a particular state with
the probability of being in that state as already mentioned in
\cite{daneshgaran_multirate}. Upon observing the basic fact that
there are four kinds of time slots, namely idle time slot,
successful transmission time slot, collision time slot, and
channel error time slot, $T_{av}$ can be evaluated by adding the
four expected slot durations:
\begin{equation} \label{eq:TAV}
T_{av} = T_I + T_C + T_S + T_E.
\end{equation}
Let us evaluate $T_I$, $T_C$, $T_S$, and $T_E$.

Upon identifying with $\sigma$ an idle slot duration, and defining
with $P_{TR}$ the probability that the channel is busy in a slot
because at least one station is transmitting:
\begin{equation}\label{eq:ptr}\small
  P_{TR} = 1 - \prod_{s = 1}^{N} (1-\tau_s)
\end{equation}
the average idle slot duration can be evaluated as follows:
\begin{equation}\small
T_I = (1-P_{TR}) \cdot \sigma
\end{equation}
The average slot duration of a successful transmission, $T_S$, can
be found upon averaging the probability $P_s^{(s)}$ that only the
$s$-th tagged station is successfully transmitting over the
channel, times the duration $T_s^{(s)}$ of a successful
transmission from the $s$-th station:
\begin{equation} \label{eq:TS}\small
T_S = \sum_{s=1}^{N} P_s^{(s)}\left(1-P_e^{(s)}\right) \cdot
T_S^{(s)}
\end{equation}
Notice that the term $(1-P_e^{(s)})$ accounts for the probability
of packet transmission without channel induced errors.

Analogously, the average duration of the slot due to erroneous
transmissions can be evaluated as follows:
\begin{equation} \label{eq:TE}\small
T_E = \sum_{s=1}^{N} P_s^{(s)} \cdot P_e^{(s)} \cdot T_E^{(s)}
\end{equation}
Let us focus on the evaluation of the expected collision slot,
$T_C$. There are $N_c$ different values of the collision
probability $P_{C}^{(d)}$, depending on the class of the tagged
station identified by $d$. We assume that in a collision of
duration $T_C^{(d)}$ (class-$d$ collisions), only the stations
belonging to the same class, or to higher classes (i.e., stations
whose channel occupancy is lower than the one of stations
belonging to the tagged station indexed by $d$) might be involved.

In order to identify the collision probability $P_{C}^{(d)}$, let us first define the following
three transmission probabilities ($P_{TR}^{C(d)},P_{TR}^{H(d)},P_{TR}^{L(d)}$) under the hypothesis
that the tagged station belongs to the class $d$. Probability $P_{TR}^{L(d)}$ represents the
probability that at least another station belonging to a lower class transmits, and it can be
evaluated as
\begin{equation}\label{eq:ptrSlow}\small
  P_{TR}^{L(d)} = 1 - \prod_{i = 1}^{d-1} \prod_{s \in n(i)} (1-\tau_s)
\end{equation}
Probability $P_{TR}^{H(d)}$ is the probability that at least one station belonging to a higher
class transmits, and it can be evaluated as
\begin{equation}\label{eq:ptrFast}\small
  P_{TR}^{H(d)} = 1 - \prod_{i = d+1}^{N_c} \prod_{s \in n(i)} (1-\tau_s)
\end{equation}
Probability $P_{TR}^{C(d)}$ represents the probability that at least a station in the same class
$d$ transmits:
\begin{equation}\label{eq:ptrClass}\small
  P_{TR}^{C(d)} = 1 - \prod_{s \in n(d)} (1-\tau_s)
\end{equation}
Therefore, the collision probability for a generic class $d$ takes
into account only collisions between at least one station of class
$d$ and at least one station within the same class (internal
collisions) or belonging to higher class (external collisions).
Hence, the total collision probability can be evaluated as:
\begin{equation}\label{eq:pColClass}\small
  P_{C}^{(d)} = P_{C}^{I(d)} + P_{C}^{E(d)}
\end{equation}
whereby
\begin{eqnarray}\label{eq:pColClass_E}\tiny
  P_{C}^{I(d)} &=& (1-P_{TR}^{H(d)}) \cdot (1-P_{TR}^{L(d)}) \cdot \\
  &&\cdot \left[P_{TR}^{C(d)} - \sum_{s \in n(d)}\tau_s\prod_{j \in n(d), j \neq\ s} (1-\tau_j)\right]\nonumber
\end{eqnarray}
represents the internal collisions between at least two stations
within the same class $d$, while the remaining are silent, and
\begin{equation}\label{eq:pColClass_H}\small
  P_{C}^{E(d)} = P_{TR}^{C(d)} \cdot P_{TR}^{H(d)} \cdot (1-P_{TR}^{L(d)})
\end{equation}
concerns to the external collisions with at least one station of
class higher than $d$.

Finally, the expected duration of a collision slot is:
\begin{equation} \label{eq:TC}\small
T_C = \sum_{d=1}^{N_c} P_C^{(d)} \cdot T_C^{(d)}
\end{equation}
Constant time durations $T_S^{(s)}$, $T_E^{(s)}$ and $T_C^{(d)}$
are defined in a manner similar to \cite{daneshgaran_multirate}
with the slight difference that the first two durations are
associated to a generic station $s$, while the latter is
associated to each duration class, which depends on the
combination of both payload length and data rate of the station of
class $d$.
\subsection{Traffic Model}
\label{subsection_traffic_model}
The employed traffic model assumes a Poisson distributed packet
arrival process, where interarrival times between two packets are
exponentially distributed with mean $1/\lambda$. In order to
greatly simplify the analysis, we consider small queue (i.e., $KQ
\rightarrow 1$, where $KQ$ is the queue size expressed in number
of packets). In order to account for the station traffic, the
Markov model employs two probabilities, $q$ and $P_{I,0}$, defined
as in \cite{daneshgaran_multirate}. Briefly, non-saturated traffic
is accounted for by a new state labelled $I$, which considers the
following two situations: 1) After a successful transmission, the
queue of the transmitting station is empty. This event occurs with
probability $(1-q^{(t)})(1-P_{eq}^{(t)})$, whereby $q^{(t)}$ is
the probability that there is at least one packet in the queue
after a successful transmission. 2) The station is in an idle
state with an empty queue until a new packet arrives in the queue.
Probability $P_{I,0}^{(t)}$ represents the probability that while
the station resides in the idle state $I$ there is at least one
packet arrival, and a new backoff procedure is scheduled. In our
analysis, each station has its own traffic; therefore, for the
$t$-th tagged station we have $q^{(t)}$ and $P^{(t)}_{I,0}$, which
can be evaluated as $q^{(t)} = 1 - e^{-\lambda^{(t)} \cdot
T_{av}}$ and $P^{(t)}_{I,0} = 1 - e^{-\lambda^{(t)} \cdot
T^{-t}_{av}}$, respectively. 
Notice that $q^{(t)}$ and $P^{(t)}_{I,0}$ stem from the fact that,
for exponentially distributed interarrival times with mean
$1/\lambda^{(t)}$, the probability of having at least one packet
arrival during time $T$ is equal to $1 - e^{-\lambda^{(t)} \cdot
T}$. Concerning $P^{(t)}_{I,0}$, the average time spent by the
system in the idle state, i.e., $T^{-t}_{av}$, is evaluated as
$T_{av}$ in (\ref{eq:TAV}) except for the fact that the tagged
station is not considered in the evaluation of the expected times
defining $T_{av}$. On the other hand, $q^{(t)}$ is calculated over
the entire average time slot duration, which also considers the
tagged station.
\section{The Proportional Fairness Throughput Allocation Algorithm}
\label{sec:optimization}
This section presents the novel throughput allocation criterion
along with a variation that proved to be useful in relation to the
packet rate of the slowest station in the network. In order to
face the fairness problem in the most general scenario, i.e.,
multirate DCF and general station loading conditions, we propose a
novel proportional fairness criterion (PFC) by starting from the
PCF defined by Kelly in \cite{kelly97}, and employed in
\cite{banchs} in connection to proportional fairness throughput
allocation in multirate and saturated DCF operations.

In the proposed model, the traffic of each station is
characterized by the packet arrival rate $\lambda_s$, which
depends mainly on the application layer. Upon setting
$\lambda_{max}$ equal to the maximum value among the packet rates
$\lambda_1,\ldots,\lambda_N$ of the $N$ contending stations,
consider the following modified PFC: 
%
\begin{equation} \label{eq:max_prop_fair_lambda}
\begin{array}{ll}
\mbox{max} & U=U(S_1,\cdots,S_N)=\sum_{s=1}^{N} \frac{\lambda_s}{\lambda_{max}} \cdot \log(S_s)\\
\mbox{subject to}&  S_s \in [0,S_{s,m}],~ s=1,\ldots,N
\end{array}
\end{equation}
whereby $S_s$ is the throughput of the $s$-th station, and $S_{s,m}$ is its maximum value, which
equal the station bit rate $R_d^{(s)}$.
%
%
In our scenario, the individual throughputs, $S_s$, are interlaced
because of the interdependence of the probabilities involved in
the transmission probabilities $\tau_s, \forall s=1,\ldots,N$. For
this reason, we reformulate the maximization problem in order to
find the $N$ optimal values of $\tau_s$ for which the cost
function in (\ref{eq:max_prop_fair_lambda}) gets maximized. Put
simply, upon starting from the optimum $\tau_s^*$, we obtain the
set of parameters of each station such that optimal point is
attained.
%

Due to 
the compactness
of the feasible region $S_s \in [0,S_{s,m}],\forall s$, the
maximum of $U(S_1,\cdots,S_N)$ can be found among the solutions of
$\nabla U=\left(\frac{\partial U}{\partial
\tau_1},\cdots,\frac{\partial U}{\partial \tau_N}\right)=0.$ After
some algebra (the proof is reported in the paper
\cite{daneshgaran_fairn_08}), the solutions can be written as:
\begin{equation} \label{eq:max_eq_tauj}
\frac{\lambda_j}{\lambda_{max}}
\frac{1}{\tau_j}-\frac{1}{1-\tau_j}\sum_{k=1, k\ne
j}^{N}\frac{\lambda_k}{\lambda_{max}}=\frac{C}{T_{av}}
\frac{\partial T_{av}}{\partial \tau_j} , \quad \forall
j=1,\ldots,N
\end{equation}
whereby $C=\sum_{i=0}^N \frac{\lambda_i}{\lambda_{max}}$, and
$T_{av}$ is a function of $\tau_1,\cdots,\tau_N$ as noted in
(\ref{eq:TAV}).

Due to the presence of $T_{av}$, a closed form of the maximum of
$U(S_1,\cdots,S_N)$ cannot be found. Notice that it is quite
difficult to derive the contribution of the partial derivative of
$T_{av}$ on $\tau_j$, especially when $N\gg 1$, because of the
huge number of network parameters belonging to different stations.
The definition of $T_{av}$ in (\ref{eq:TAV}) is composed by four
different terms, anyone of which includes the whole set of
$\tau_s$, $\forall s$. In order to overcome this problem, we first
obtain the optimal values $\tau_s^*, \forall s$ from
(\ref{eq:max_eq_tauj}) by means of Mathematica. Then, we choose
the value of the minimum contention window size, $W^{(s)}_0$, by
equating the optimizing $\tau_s^*$ to (\ref{eq:tau_s}) for any
$s$.

The results of the optimization problem
(\ref{eq:max_prop_fair_lambda}) will be denoted by the acronym LPF
in the following.

Let us derive some observations on the proposed throughput
allocation algorithm by contrasting it to the classical PF
algorithm. Upon employing the classical PF method, a throughput
allocation is proportionally fair if a reduction of $x\%$ of the
throughput allocated to one station is counterbalanced by an
increase of more than $x\%$ of the throughputs allocated to the
other contending stations. The key observation in our setup can be
summarized as follows. Consider the two stations above with packet
rates $\lambda_1=50~ \textrm{pkt/s}$ and $\lambda_2=100~
\textrm{pkt/s}$, respectively. The ratio $\lambda_1/\lambda_2$ can
be interpreted as the frequency by which the first station tries
to get access to the channel relative to the other station. By
doing so, in our setup a throughput allocation is proportionally
fair if, for instance, a reduction of $20\%$ of the throughput
allocated to the first station, which has a relative frequency of
$1/2$, is counterbalanced by an increase of more than $40\%$ of
the throughput allocated to the second station. In a scenario with
multiple contending stations, the relative frequency is evaluated
with respect to the station with the highest packet rate in the
network, which gets unitary relative frequency.

Based on extensive analysis, we found that the optimization
problem (\ref{eq:max_prop_fair_lambda}) sometimes yields
throughput allocations that cannot be actually managed by the
stations. As a reference example, assume that, due to the specific
channel conditions experienced, the first station has a bit rate
equal to $1$ Mbps and needs to transmits $200~ \textrm{pkt/s}$.
Given a packet size of $1024$ bytes, that is $8192$ bits, the
first station would need to transmit $8192\times 200~
\textrm{bps}\approx 1.64 \textrm{Mbps}$ far above the maximum bit
rate decided at the physical layer. In this scenario, such a
station could not send over the channel a throughput greater than
1Mbps. The same applies to the other contending stations in the
network experiencing similar conditions. In order to face this
issue, we considered the following optimization problem
\begin{equation} \label{eq:max_prop_fair_lambda_trunc}
\begin{array}{ll}
\mbox{max} & \sum_{s=1}^{N} \frac{\lambda^*_s}{\lambda^*_{max}} \cdot \log(S_s)\\
\mbox{over}&  S_s \in [0,S_{s,m}],~ s=1,\ldots,N
\end{array}
\end{equation}
whereby, $\forall s=1,\ldots,N$, it is
\[
\lambda^*_s=\left\{
\begin{array}{ll}
\lambda_s & \textrm{if} ~\lambda_s\cdot PL^{(s)}\cdot 8 \le R_d^{(s)}\\
 \frac{R_d^{(s)}}{8\cdot PL^{(s)}}           & \textrm{if} ~\lambda_s\cdot PL^{(s)}\cdot 8 > R_d^{(s)}
\end{array} \right.
\]
and $\lambda^*_{max}=\max_{s}\lambda^*_s$. The allocation problem
in (\ref{eq:max_prop_fair_lambda_trunc}), solved as for the LPF in
(\ref{eq:max_prop_fair_lambda}), guarantees a throughput
allocation which is proportional to the frequency of channel
access of each station relative to their actual ability in
managing such traffic. The results of the optimization problem
(\ref{eq:max_prop_fair_lambda_trunc}) will be denoted by the
acronym MLPF in the following section.
\section{Simulation Results}
\label{SimulationResults_Section}
\figuramedia{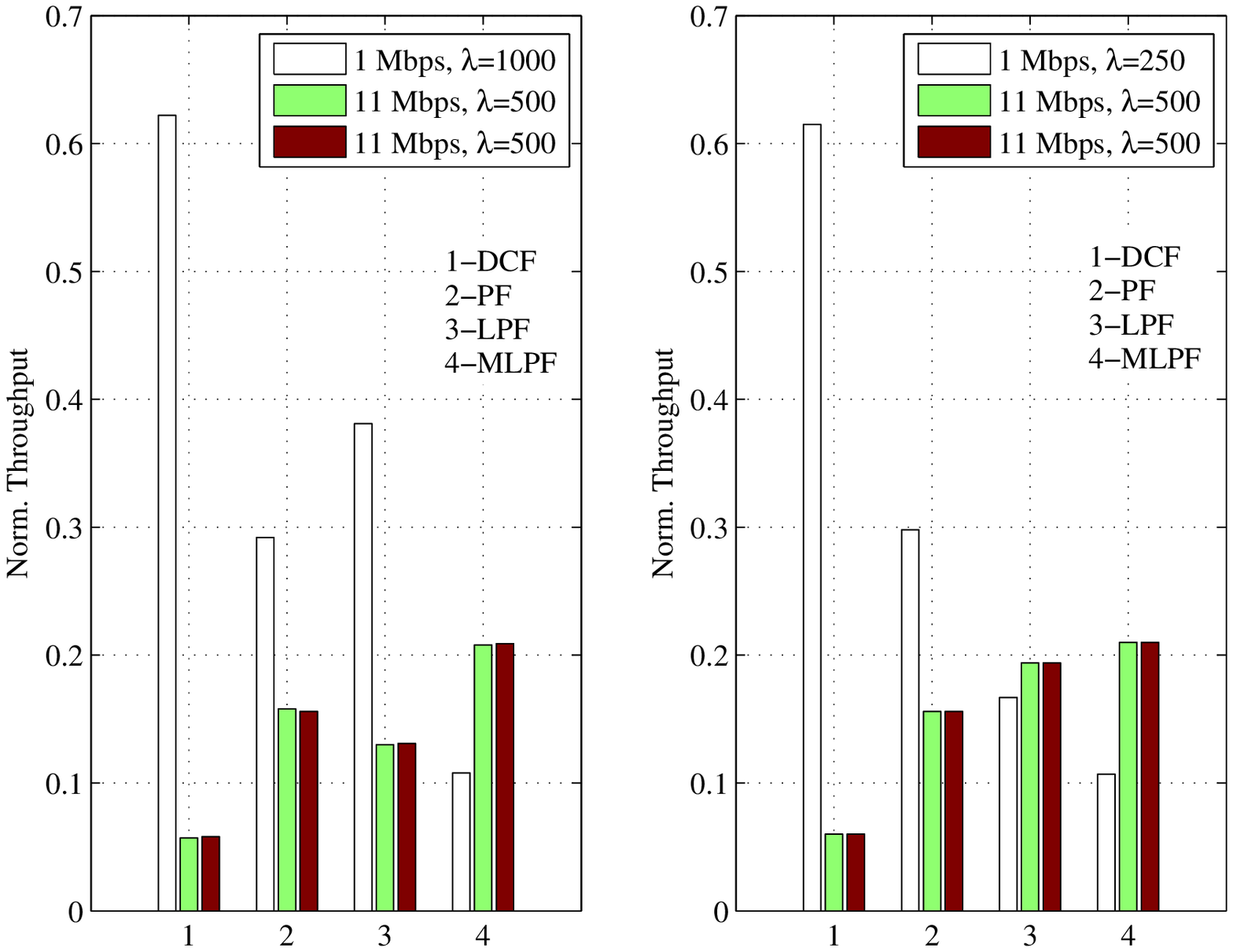}{Simulated normalized throughput
achieved by three contending stations upon employing 1) a
classical DCF; 2) DCF with PF allocation; 3) DCF optimized as
noted in (\ref{eq:max_prop_fair_lambda}); and 4) DCF optimized
with the criterion MLPF. Left and right plots refer to scenarios A
and B, respectively.}{fig_PF_1}
This section presents some preliminary simulation results obtained
for two network scenarios optimized with the fairness criteria
proposed in the previous section. Typical MAC layer parameters for
IEEE802.11b~\cite{standard_DCF_MAC} have been used for performance
validation. Due to space limitations, we invite the interested
reader to consult \cite{daneshgaran_multirate} for a list of the
network parameters employed here as well as for the details about
the employed simulator. Moreover, further results on the behaviour
of the proposed criteria are available on the work
\cite{daneshgaran_fairn_08}.

The first investigated scenario, namely A, considers a network with $3$ contending stations. Two
stations transmit packets with rate $\lambda=500~\textrm{ pkt/s}$ at $11$ Mbps. The payload size,
assumed to be common to all the stations, is $PL=1028$ bytes. The third station has a bit rate
equal to 1Mbps and a packet rate $\lambda=1000~\textrm{ pkt/s}$. The simulated normalized
throughput achieved by each station in this scenario is depicted in the left subplot of
Fig.~\ref{fig_PF_1} for the following four setups. The three bars labelled 1-DCF represent the
normalized throughput achieved by the three stations with a classical DCF. The second set of bars,
labelled 2-PF, identifies the simulated normalized throughput achieved by the DCF optimized with
the PF criterion \cite{banchs,kelly97}, whereby the actual packet rates of the stations are not
considered. The third set of bars, labelled 3-LPF, represents the normalized throughput achieved by
the three stations when the allocation problem (\ref{eq:max_prop_fair_lambda}) is employed.
Finally, the last set of bars, labelled 4-MLPF, represents the simulated normalized throughput
achieved by the contending stations when the CW sizes are optimized with the modified fairness
criterion. Notice that the throughput allocations guaranteed by LPF and MLPF improve over the
classical DCF. When the station packet rate is considered in the optimization framework, a higher
throughput is allocated to the first station presenting the maximum value of $\lambda$ among the
considered stations. However, the highest aggregate throughput is achieved when the allocation is
accomplished with the optimization framework 4-MLPF. The reason for this behaviour is related to
the fact that the first station requires a traffic equal to $8.22\textrm{ Mbps}=10^3 \textrm{
pkt/s}\cdot 1028 \textrm{ bytes/pkt}\cdot 8 \textrm{ bits/pkt}$, which is far above the maximum
traffic ($1\textrm{ Mbps}$) that the station would be able to deal with in the best scenario. In
this respect, the criterion MLPF results in better throughput allocations since it accounts for the
real traffic that the contending station would be able to deal with in the specific scenario at
hand.

Similar considerations can be drawn from the results shown in the
right subplot of Fig.~\ref{fig_PF_1} (related to scenario B),
whereby in the simulated scenario the two fastest stations are
also characterized by a packet rate greater than the one of the
slowest station. Notice that the optimization framework 3-LPF is
able to guarantee improved aggregate throughput with respect to
both the non-optimized DCF and the classical PF algorithms.

The aggregate throughput achieved in the two investigated
scenarios are noted in Table \ref{JainsFairnessIndex}, whereby we
also show the fairness Jain's index \cite{jain} evaluated on the
normalized throughputs noted in the subplots of
Fig.~\ref{fig_PF_1}. It is worth noticing that the proposed
throughput allocation criteria are able to guarantee either
improved aggregate throughput, and improved fairness among the
contending stations over both the classical DCF and the PF
algorithm. Moreover, notice that the fairness index and the
aggregate throughput of both DCF and PF do not change in the two
scenarios, since they do not account for the actual packet rates
that the contending stations need to transmit over the channel.
\figura{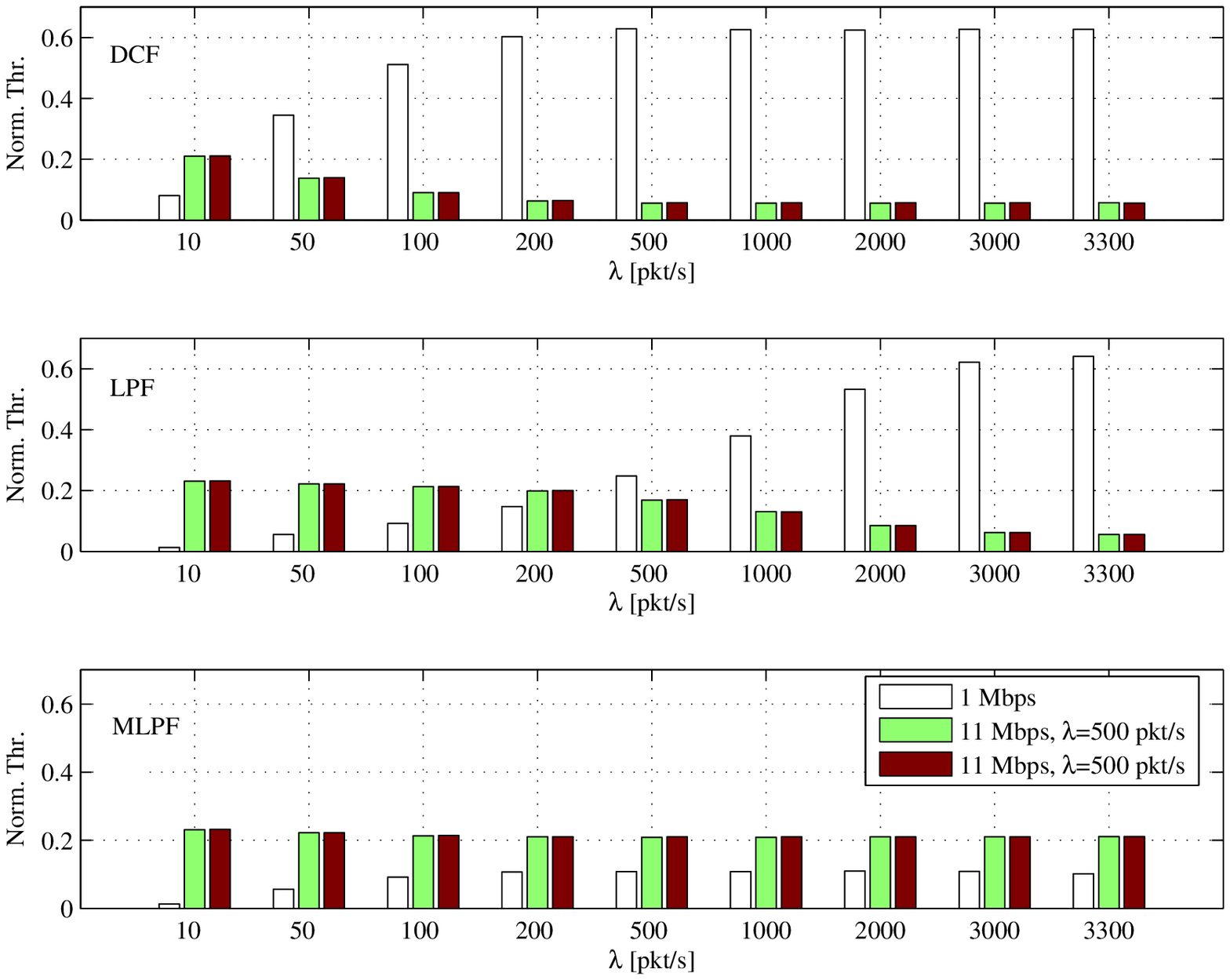}{Simulated normalized throughput
achieved by three contending stations as a function of the packet
rate of the slowest station in DCF, LPF and MLPF
modes.}{ThroughDCF_simfin}

For the sake of investigating the behaviour of the proposed
allocation criteria as a function of the packet rate of the
slowest station, we simulated the throughput allocated to a
network composed by three stations, whereby the slowest station,
transmitting at 1Mbps, presents an increasing packet rate in the
range 10$-$3300 pkt/s. The other two stations transmit packets at
the constant rate $\lambda=500$ pkt/s at 11 Mbps. The simulated
throughput of the three contending stations is shown in the three
subplot of Fig. \ref{ThroughDCF_simfin} for the unoptimized DCF,
as well as for the two criteria LPF and MLPF. Some considerations
are in order. Let us focus on the throughput of the DCF (uppermost
subplot in Fig. \ref{ThroughDCF_simfin}). As far as the packet
rate of the slowest station increases, the throughput allocated to
the fastest stations decreases quite fast because of the
performance anomaly of the DCF \cite{heusse}. The three stations
reach the same throughput when the slowest station presents a
packet rate equal to $500 \textrm{pkt/s}$, corresponding to the
one of the other two stations. From $\lambda=500 \textrm{pkt/s}$
all the way up to 3300 pkt/s, the throughput of the three stations
do not change anymore, since all the stations have a throughput
imposed by the slowest station in the network. Let us focus on the
results shown in the other two subplots of Fig.
\ref{ThroughDCF_simfin}, labelled LPF and MLPF, respectively. A
quick comparison among these three subplots in Fig.
\ref{ThroughDCF_simfin} reveals that the allocation criterion MLPF
guarantees improved aggregate throughput for a wide range of
packet rates of the slowest station, greatly mitigating the rate
anomaly problem of the classical DCF operating in a multirate
setting. In terms of aggregate throughput, the best solution is
achieved with the criterion MLPF.
\begin{table}\caption{Jain's Fairness Index and Aggregate Throughput S}
\small
\begin{center}
\begin{tabular}{l|l||c|c|c|c}\hline
\hline
\multicolumn{2}{c||}{Scenarios in Fig.~\ref{fig_PF_1}} & 1-DCF & 2-PF & 3-LPF & 4-MLPF \\
\hline
\multirow{2}{*}{A} & Jain's Index   &  0.460 & 0.909  & 0.766  & 0.9317   \\
                   & S [Mbps]     &  1.89  & 3.74   & 3.25 & 4.69 \\
\hline
\multirow{2}{*}{B} & Jain's Index & 0.467  & 0.902  & 0.995  & 0.9290  \\
                   & S [Mbps]     &  1.93  & 3.73   & 4.43 & 4.72\\

\hline\hline
\end{tabular}
 \label{JainsFairnessIndex}
\end{center}
\end{table}
\end{document}